\documentclass[aps,prx,twocolumn,groupaddress,showpacs]{revtex4-1}
\usepackage{amsmath,amssymb,graphicx,stmaryrd}
\usepackage{physics}
\usepackage[utf8]{inputenc}
\usepackage[T1]{fontenc}
\usepackage{xcolor}

\IfFileExists{newtxtext.sty}
   {\usepackage{newtxtext,newtxmath}}
   {\IfFileExists{stix.sty}
      {\usepackage{stix}}
      {\IfFileExists{mathptmx.sty}
      {\usepackage{mathptmx}}{} } }

\usepackage{booktabs,siunitx}
\newcommand{\valunit}[2]{\SI[inter-unit-product =\ensuremath{\!\cdot\!},per-mode=reciprocal]{#1}{#2}}

\pdfoutput=1
\definecolor{LinkColor}{rgb}{0.1,0.2,0.4}
\usepackage{hyperref}
\hypersetup{
   pdfauthor={Sudeep Adhikari and K. S. D. Beach},
   pdftitle={Reliable extraction of energy landscape properties from critical force distributions},
   pdfsubject={Escape rate from a confining well},
   pdfkeywords={single-molecule pulling experiments,} {optical tweezers,} {energy landscape,} {escape rate,} {rupture force distribution,} {Langevin dynamics,} {diffusion},
   colorlinks=true,
   citecolor=LinkColor,
   linkcolor=LinkColor,
   urlcolor=LinkColor
   }

\DeclareMathOperator{\ei}{Ei}
\newcommand{\linkdoi}[2]{\href{http://dx.doi.org/#2}{#1}}
\newcommand{\linkisbn}[2]{\href{https://isbndb.com/book/#2}{#1}}

\begin{document}

\title{Reliable extraction of energy landscape properties from critical force distributions}

\author{Sudeep Adhikari}
\email[Electronic mail:\ ]{sadhika6@go.olemiss.edu}
\affiliation{Department of Physics and Astronomy, The University of Mississippi, University, Mississippi 38677, USA}

\author{K. S. D. Beach}
\email[Electronic mail:\ ]{kbeach@olemiss.edu}
\affiliation{Department of Physics and Astronomy, The University of Mississippi, University, Mississippi 38677, USA}

\date{May 15, 2020}

\begin{abstract}
The structural dynamics of a biopolymer is governed by a process of diffusion through its conformational energy landscape. In pulling experiments using optical tweezers, features of the energy landscape can be extracted from the probability distribution of the critical force at which the polymer unfolds. The analysis is often based on rate equations having Bell-Evans form, although it is understood that this modeling is inadequate and leads to unreliable landscape parameters in many common situations. Dudko and co-workers [\linkdoi{Phys.\ Rev.\ Lett.\ {\bf 96}, 108101 (2006)}{10.1103/PhysRevLett.96.108101}] have emphasized this critique and proposed an alternative form that includes an additional shape parameter (and that reduces to Bell-Evans as a special case). Their fitting function, however, is pathological in the tail end of the pulling force distribution, which presents problems of its own. We propose a modified closed-form expression for the distribution of critical forces that correctly incorporates the next-order correction in pulling force and is everywhere well behaved. Our claim is that this new expression provides superior parameter extraction and is valid even up to intermediate pulling rates. We present results based on simulated data that confirm its utility.
\end{abstract}

\maketitle
\section{ \label{SEC:introduction} Introduction }
The contribution of explicitly quantum processes notwithstanding~\cite{Stohr-SciAdv-19}, 
classical energy landscape theory~\cite{Bryngelson-PNAS-87, Onuchic-AnnRevPhysChem-97, Galzitskaya-PNAS-99, Onuchic-AdvProtChem-00} provides a useful framework for describing the evolution of biopolymers between various folded and unfolded configurations through a process of thermally driven escape from local confining potentials~\cite{Talkner-PLA-82}. Developing tools of analysis within this framework has become ever more pressing, given the profound developments in single-molecule biophysics~\cite{Merkel-Nat-99,Liphardt-Sci-01,Li-JCP-04,Kirmizialtin-JCP-05,Hinterderfer-NatMethods-06,Gilbert-NanoLett-07,Greenleaf-ARBBS-07,Neupane-NAR-11,Souza-NatMethods-12,Bull-ACSNano-14,Edwards-NanoLett-15,Patten-CPC-17,Okoniewski-NAR-17,Yu-Science-17,Walder-ACSNano-18,Walder-NanoLett-18}. One of the key practical problems is how to infer the energy landscape, or at least a projection of it onto an appropriate reaction coordinate, from experimentally measured quantities~\cite{Jarzynski-PRL-97,Hummer-PNAS-01,Harris-PRL-07,Gupta-NatPhys-11,Zhang-JSP-11,Engel-PRL-14,Manuel-PNAS-15,Heenan-JCP-18,Alamilla-JTB-19,Alamilla-PRE-19}. As is typical of inverse problems, recovery of the landscape from measured data is ill conditioned: it is highly sensitive to experimental uncertainties and to any assumptions that go into the forward model.

In pulling experiments using optical tweezers~\cite{Litvinov-PNAS-02}, the determination of landscape features has historically been carried out based on Bell-Evans phenomenological theory~\cite{Bell-Sci-78,Evans-BPJ-91,Evans-BPJ97,Rief-Sci-97,Rief-PRL-98}, which assumes that the rate constant $k(F)$ scales up exponentially with applied force from its unperturbed, intrinsic value $k_0$ according to the Arrhenius law,
\begin{equation} \label{EQ:Bell-Evans}
 k_\text{BE}(F) = k_0 e^{\,\beta F x^\ddagger}\!.
\end{equation}
Here, $\beta ^{-1} = k_BT$ is the thermal energy scale set by the aqueous environment; $x^\ddagger$ is the minimum-to-barrier distance of the effective one-dimensional potential $U_0(x)$, a continuous (but not necessarily smooth) function of the end-to-end extension.

A common experimental situation involves the application of a pulling force $F = KVt$ that grows linearly in time until the rupture force $F_c$ is reached. Although other pulling protocols are sometimes employed~\cite{Woodside-PNA-06,Woodside-AR-14, Barsegov-PRL-05, Maitra-PRL-10, Ritchie-COSB-15}, we focus on the case of constant pulling speed, and we ignore instrument-specific issues of compliance~\cite{Woodside-BPJ-14,Makarov-JCP-14,Nam-JCP-14}.

It is recognized that a description of pulling experiments based on the Bell-Evans formula for the force-induced rupture rate is in poor accord with results from numerical simulations~\cite{Hummer-BPJ-03}. The naive thermal-activation picture, represented by the Bell-Evans theory, suffers from various inadequacies that are important to address. To begin, Eq.~\eqref{EQ:Bell-Evans} is strictly applicable only in the limit of low pulling rate ($KV \lesssim KV_\text{min} = k_0/\beta x^\ddagger$) and ultrahigh barrier ($\Delta G^\ddagger \gg Fx^\ddagger, k_BT$). Even in the moderate pulling regime, it incorrectly predicts the rupture force distribution. It also ignores self-consistency effects in the sense that it does not account for the fact that the distance $x^\ddagger$ and the energy barrier ${\Delta G}^\ddagger$ are themselves force dependent and both diminish with increasing $F$ as the energy landscape is tilted. Nor does it properly account for the shape of the barrier, which plays a vital role in establishing the escape rate and the nature of the escape trajectory for more modest, biologically relevant barrier heights.

Consequently, there are many situations in which the phenomenological theory incorrectly predicts the results of pulling experiments. It tends to overestimate the rate of rupture, $k(F)$, at a given force $F$ and to underestimate the mean and most-probable rupture forces. Hence, when the Bell-Evans rate, $k_\text{BE}(F)$, is used as the basis for a fit to experimental data, the extracted parameters, ${\Delta G}^\ddagger$, $x^\ddagger$, and $k_0$, may be incorrectly predicted. Our main concern here lies in the reliable extraction of these physical quantities.

Attempts have been made to improve on the Bell-Evans theory by introducing additional fitting parameters~\cite{Garg-PRB-95,Friddle-PRL-08}, sometimes in an {\it ad hoc} way. Dudko and co-workers have tried to make the analysis more rigorous~\cite{Dudko-PRL-06}. They calculated $k(F)$ and the corresponding probability density of the rupture force $p(F_c)$ within the framework of Kramer's theory~\cite{Kramers-Phy-40} for two specific free energy surfaces---the cusp surface and the linear cubic surface---and showed that these two examples can be subsumed into a single result [appearing as Eq.~(3) in Ref.~\onlinecite{Dudko-PRL-06}], 
\begin{equation}\label{EQ:rate-Dudko}
k_{\text{D}}(F) = k_0 \biggl(1-\frac{\nu F x^\ddagger}{\Delta G^\ddagger}\biggr)^{1/\nu-1}
e^{\beta \Delta G^\ddagger\bigl[1-(1-\nu F x^\ddagger/\Delta G^\ddagger)^{1/\nu} \bigr]}, 
\end{equation}
with interpolation provided by a shape parameter $\nu$. This encompasses the Bell-Evans result, since $k_{\text{D}}(F) \to k_{\text{BE}}(F)$ as $\nu \to 1$. It is clear, however, that for all $\nu \neq 1$ Eq.~\eqref{EQ:rate-Dudko} has a dangerous point of nonanalyticity. The vanishing of the rate $k_\text{D}(F) \to 0$ as $F \to \Delta G^\ddagger/x^\ddagger\nu$ (for shape parameters in the range $0 < \nu < 1$) is manifestly unphysical; hence the Dudko expression is only appropriate for the pulling regime in which $F \ll \Delta G^\ddagger/x^\ddagger\nu$. In fact, the region of validity is more constrained still, since we should further require that the escape rate grow with pulling force. As it turns out, the function $k_{\text{D}}(F)$ is monotonic increasing only for
\begin{equation} 
F < \frac{\Delta G^\ddagger}{x^\ddagger\nu}\Biggl[1 - \biggl(\frac{1-\nu}{\beta\Delta G^\ddagger}\biggr)^{\nu}\Biggr].
\end{equation}

We pursue a different approach that produces no nonanalyticity and no obviously unphysical behavior. We compute $\log k(F)/k_0$ order by order in the pulling force. Rather than truncate the expansion, we approximate the higher-order terms as a resummation by geometric series---similar in spirit to the random phase approximation or the infinite summation of ladder diagrams in many-body theory:
\begin{equation} \label{EQ:rate-resummed}
k(F) = k_0 \exp\biggl(\frac{\beta Fx^{\ddagger}}{1+F/2 \kappa^{\ddagger} x^{\ddagger}}\biggr).
\end{equation}
Here $\kappa^\ddagger$ is the reduced curvature of the well and barrier. The route to Eq.~\eqref{EQ:rate-resummed} is nothing more than a mathematical trick, but it rather elegantly cures the ill behavior of a truncated expansion, and it fortuitously leads to a closed-form expression for the cumulative probability distribution.

Our attempts to benchmark Eq.~\eqref{EQ:rate-resummed} fall into two categories: prediction and parameter extraction, which correspond to the forward and inverse problems. In the forward direction, we determine the escape rates and the cumulative probability distribution of the critical force following the numerical method described in Sec.~\ref{SEC:Numerical-Simulations}. We compare the simulated behavior to the various analytical predictions. We find that our proposal outperforms the Bell-Evans and Dudko expressions, across many different choices of energy landscape and over a broad range of pulling rates. In the inverse direction, analytical forms for the cumulative probability distribution $P(F_c)$ are fit to the simulated data to extract the optimal values of the intrinsic parameters $k_0$, $x^\ddagger$, and $\kappa^\ddagger$. 

The results we achieve are compelling. The values of the three parameters that we extract are in excellent agreement with the actual values that characterize the underlying energy landscape. Moreover, the agreement appears to hold over an unexpectedly large range of pulling rates, with $KV/KV_\text{min}$ spanning six or seven orders of magnitude.

In contrast, fits of simulation data to the Bell-Evans cumulative probability distribution, insofar as they are able to produce good values of $k_0$ and $x^\ddagger$ at all, only do so at the very slowest pulling rates. It is difficult to speak definitively of how well Dudko's expression performs, since in that context fits must be carried out in conjunction with a force cutoff somewhere below the point of nonanalyticity. This is an unwelcome complication. The cutoff itself introduces a significant element of uncertainty in the fit, since where best to put the cutoff cannot be determined if the landscape is not yet known.

\section{ \label{SEC:Formal-Development} Formal Development}

\begin{figure}
\begin{center}
\includegraphics[width=3.0in]{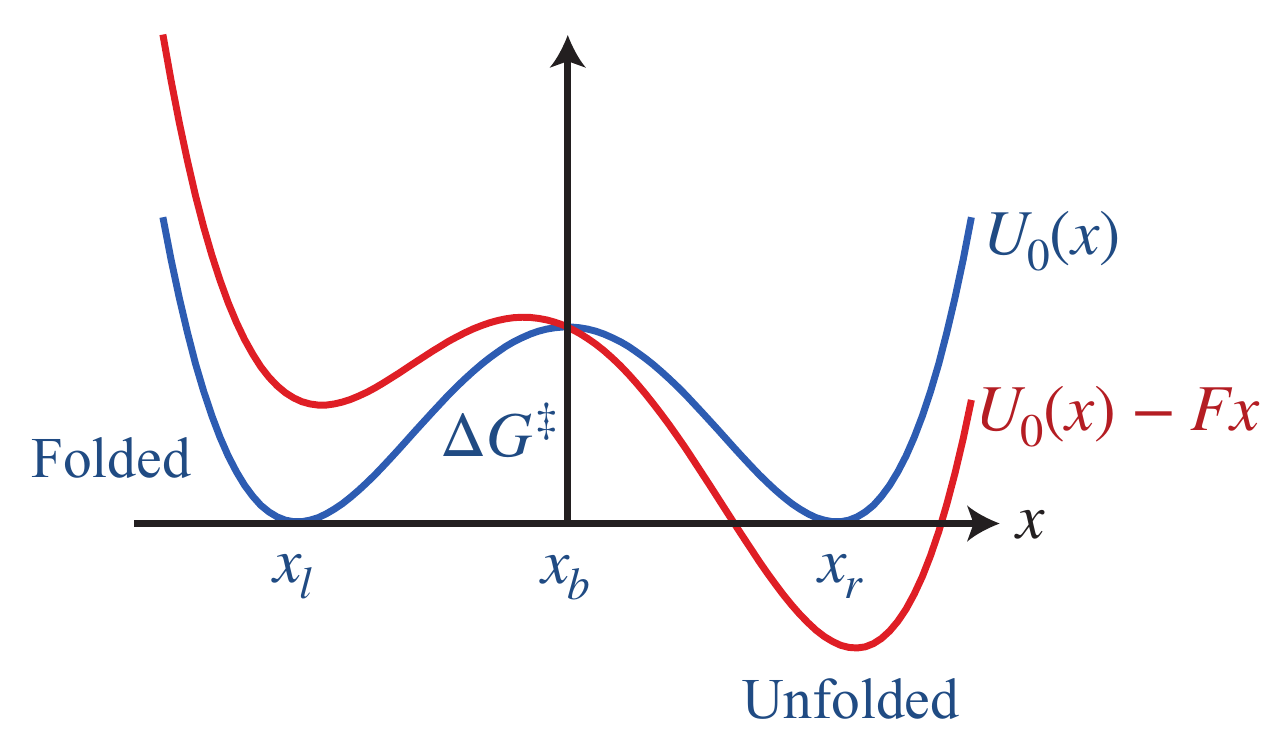}
\end{center}
\caption{\label{FIG:Double-well} 
Blue curve: The double-well potential has equilibrium positions at $x_l$ and $x_r$, separated by a barrier at $x_b$. A  particle escaping from left to right experiences a barrier of height ${\Delta G}^\ddagger = U_0(x_b) - U_0(x_l)$, peaked at a distance $x^\ddagger = x_b- x_l$ from the bottom of the left well. Red curve: After application of a pulling force $F$, the energy landscape has tilted to favor the destination well on the right. Observe that the well positions have shifted and that the height of the barrier holding the particle in the left well has decreased.}
\end{figure}

Kramers theory tells us that the escape rate depends weakly (polynomially) on the curvature at the bottom of the well and the top of the barrier but strongly (exponentially) on the height of the apparent energy barrier in the direction of travel~\cite{Kramers-Phy-40,Hanggi-RMP-80}. We consider a double well potential $U_0(x)$, with wells at positions $x_l$ and $x_r$ separated by a barrier at $x_b$ ($x_l < x_b < x_r$), as illustrated in Fig.~\ref{FIG:Double-well}. The well escape rate from left to right is given by $k_0 \sim \exp(-\beta \Delta U_0)$, where $\Delta U_0 = U_0(x_b) - U_0(x_l)$.

We allow for a pulling force $F$ that tilts the potential landscape according to
\begin{equation}\label{EQ:Pot-landscape}
U(x) = U_0(x) - Fx.
\end{equation}
The corresponding rate equation becomes
\begin{equation}
k(F)\sim \exp\bigl[-\beta  \bigl(U(x_b + \delta x_b) - U(x_l + \delta x_l )\bigr)\bigr],
\end{equation}
where $\delta x_l$ and $\delta x_r$ denote the shifts in the well positions as a result of the tilt. Taylor expansions of the extremal conditions $U'(x_l + \delta x_l) = 0 $ and $U'(x_b + \delta x_b) = 0 $ around $x_l $ and $x_b$ up to first order in $\delta x_l$ and $\delta x_b$ give $\delta x_l = F/U_0''(x_l) = F/\kappa_l$ and $\delta x_b = F/U_0''(x_b) = -F/\kappa_b$. A further expansion of  $U(x_b + \delta x_b)$ and $U(x_l + \delta x_l)$ around $x_b$ and $x_l$, respectively, up to second order in $F$, yields a rate equation of the form
\begin{equation} \label{EQ:rate-truncated}
k(F)= k_0 \exp\biggl[\beta Fx^{\ddagger}\biggl(1-\frac{F}{2 \kappa^{\ddagger} x^{\ddagger}}\biggr)\biggr].
\end{equation}
Here, $x^\ddagger = x_b - x_l$, and
\begin{equation}
\frac{1}{\kappa^\ddagger} = \frac{1}{U_0''(x_l)} - \frac{1}{U_0''(x_b)} = \frac{1}{\kappa_l} + \frac{1}{\kappa_b}.
\end{equation}

\begin{figure}
\begin{center}
\includegraphics[width=3.0in]{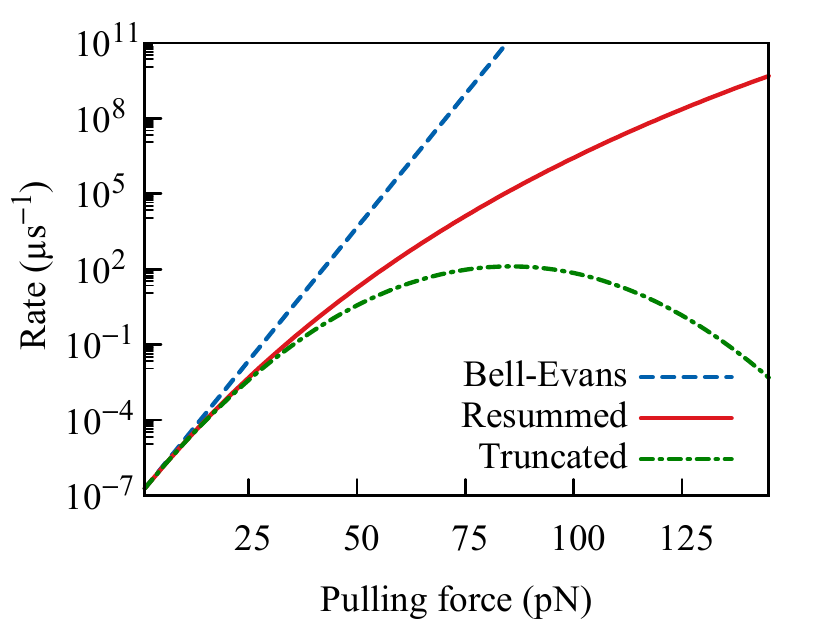}
\end{center}
\caption{\label{FIG:rate-comp} 
The well escape rate $k(F)$ is plotted against the applied pulling force $F$. The upper (dashed blue) curve corresponds to the Bell-Evans (BE) rate [Eq.~\eqref{EQ:Bell-Evans}], the middle (solid red) to our proposed infinite-resummation expression [Eq.~\eqref{EQ:rate-resummed}], and the lower (dot-dashed green) to an expansion truncated at second order in the pulling force [Eq.~\eqref{EQ:rate-truncated}]. The BE result grows exponentially without bound (but shows as a straight line because of the log-linear scale). The truncated expression turns over and becomes unphysical around \SI{80}{\pico\newton}. The resummed form strikes a middle course, growing monotonically but saturating at a large, finite value, $k_0\exp\bigl[2\beta \kappa^\ddagger (x^\ddagger)^2\bigr]$.
}
\end{figure}

Successive terms in the expansion of $\log k(F)/k_0$ have alternating sign, which is important for proper convergence of the series. Indeed, there is no polynomial expression, arising as a truncation of the series at finite order, that does not either substantially over- or undershoot the true rate for large applied $F$. The negative-prefactor terms at even powers of $F$ are particularly troublesome, because they lead to nonmonotonicity. As a workaround, we make use of the idea of infinite resummation, $ 1- \epsilon + \epsilon^2 - \cdots \approx 1/(1+ \epsilon)$, which transforms Eq.~\eqref{EQ:rate-truncated} into Eq.~\eqref{EQ:rate-resummed}, at least up to discrepancies at $O(F^3)$. The transformed expression is well behaved everywhere and displays no obviously unphysical behavior (see Fig.~\ref{FIG:rate-comp}). Moreover, it leads to a closed-form expression for the cumulative probability distribution (with the correct normalization $P(F_c)\rightarrow 1$ as $F_c\rightarrow\infty$; Dudko's expression, in contrast, cannot be properly normalized). 

In the usual adiabatic limit, the expression for the cumulative probability distribution of the rupture force is given by 
\begin{equation} \label{EQ:cum-prob}
P(F_c)= 1- \exp\Biggl[-\int_{0}^{F_c}\!\frac {dF}{\dot F} k(F)\Biggr].
\end{equation}
Equations~\eqref{EQ:Bell-Evans} and \eqref{EQ:cum-prob} together give the cumulative probability distribution of the rupture force as predicted by the Bell-Evans phenomenological model,
\begin{equation} \label{EQ:cum-prob-BE}
P_{\text{BE}}(F_c)= 1- \exp\biggl[\frac{k_0}{KV\beta x^{\ddagger}}\bigl(1- e^{\,\beta F_c x^{\ddagger}}\bigr)\biggr].
\end{equation}
If instead we put Eq.~\eqref{EQ:rate-resummed} into Eq.~\eqref{EQ:cum-prob}, we get a more complicated result, but one that is still simple enough to use for fitting (e.g., via the Marquardt-Levenberg method):
\begin{equation} \label{EQ:cum-prob-resummed}
P(F_c)= 1- \exp\biggl[-\frac{k_0}{KV}\bigl(F_1 + F_2 - 2 x^{\ddagger} \kappa^{\ddagger}\bigr)\biggr].
\end{equation}
The quantities $F_1$ and $F_2$ have units of force and are explicit functions of the critical value $F_c$:
\begin{equation} \label{EQ:F1-F2-definitions}
\begin{split}
F_1 &= \bigl(F_c + 2x^{\ddagger} \kappa^{\ddagger}\bigr)\exp\Biggl(\frac {2F_c{x^{\ddagger}}^2 \beta {\kappa}^{\ddagger}}{F_c+ 2 x^{\ddagger}{\kappa}^{\ddagger}}\Biggr),\\
F_2 &= 4 {x^{\ddagger}}^3 \beta {{\kappa}^{\ddagger}}^2 \exp \bigl( 2{x^{\ddagger}}^2 \beta {\kappa}^{\ddagger}\bigr)\\
 &\qquad \times \Biggl[\ei\Biggl(-\frac{4 {x^{\ddagger}}^3 \beta {{\kappa}^{\ddagger}}^2}{F_c + 2x^{\ddagger} \kappa^{\ddagger}}\Biggr) - \ei\bigl(-2{x^{\ddagger}}^2 \beta {\kappa}^{\ddagger}\bigr)\Biggr].
\end{split}
\end{equation}
The exponential integral $\ei(x)= -\int_{-x}^{\infty}dt\,t^{-1}e^{-t}$ is a standard special function that is available in most data analysis software.

\begin{figure*}
\begin{center}
\includegraphics[width=2in]{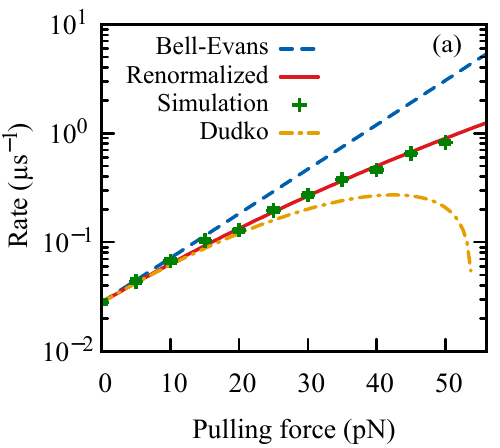}
\includegraphics[width=2in]{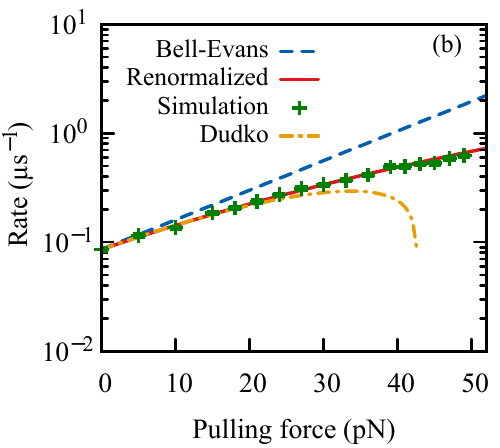}
\includegraphics[width=2in]{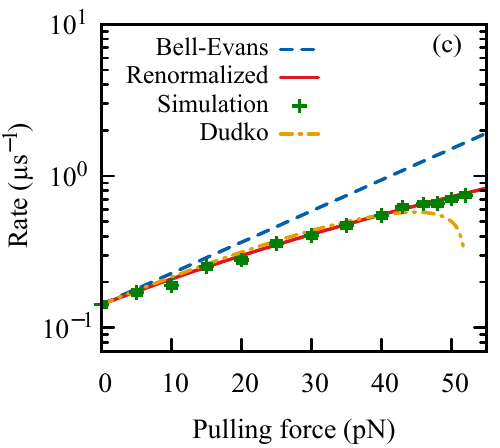}
\includegraphics[width=2in]{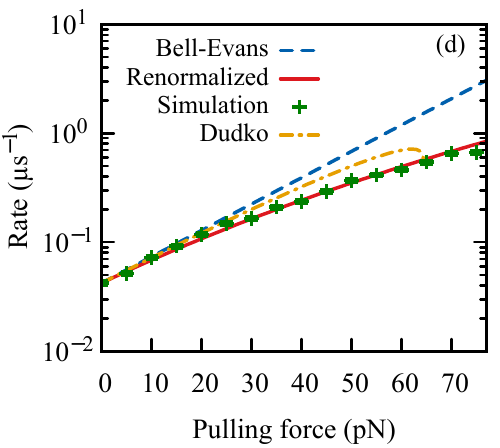}
\includegraphics[width=2in]{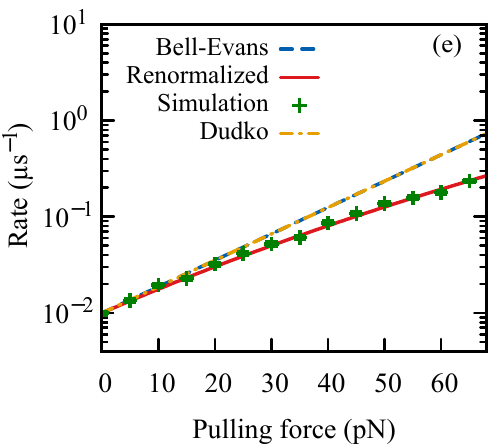}
\includegraphics[width=2in]{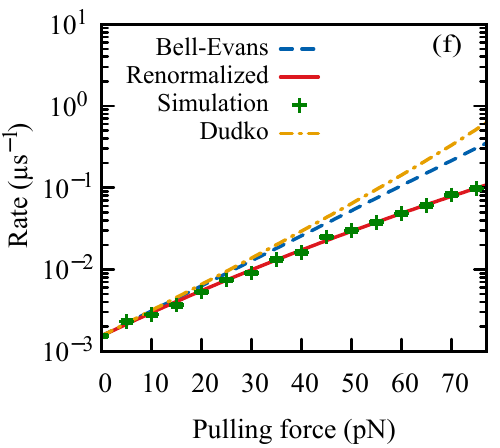}
\caption{\label{FIG:nu}
Numerical measurements of the escape rate (green data points with error bars) are plotted versus the applied pulling force. Also shown for comparison are the predictions of the Bell-Evans approach [Eq.~\eqref{EQ:Bell-Evans}, blue dashed line], Dudko approach [Eq.~\eqref{EQ:rate-Dudko}, dot-dashed orange line], and our renormalized rate equation [Eq.~\eqref{EQ:rate-renorm} with a common value $\alpha = 1/3$, solid red line]. The simulations were carried out for potentials with various values of $\nu = 2\Delta G^\ddagger/ \kappa^\ddagger (x^\ddagger)^2$, the shape parameter: (a)~$\nu  = 0.66$, (b)~$\nu = 0.75$, (c)~$\nu = 0.82$, (d)~$\nu = 0.9$, (e)~$\nu = 1.0$, and (f)~$\nu = 1.1$. We note the remarkable agreement between simulation and the renormalized form. No fitting is involved.} 
\end{center}
\end{figure*}

The choice $F = KVt$ is helpful here but not essential. Its main advantage is that the
differential appearing in Eq.~\eqref{EQ:cum-prob} simplifies to $dF/\dot{F} = (KV)^{-1}dF$, and hence the integration measure is trivial. The closed-form expression that we obtain in Eqs.~\eqref{EQ:cum-prob-resummed} and \eqref{EQ:F1-F2-definitions} does depend on this choice. But any pulling schedule $F(t)$ that is monotonic increasing (so that $\dot{F}$ never vanishes or goes negative) and growing at most as a polynomial in $t$ can be treated similarly.

We now comment on the connection to the prior work of Dudko and co-workers. The unperturbed potential $U_0(x)$ can be expanded to quadratic order around the bottom of the well, $U_{0,l}(x) = U_0(x_l) + (\kappa_l/2)(x-x_l)^2$, and around the peak of the barrier, $U_{0,b}(x) = U_0(x_b) - (\kappa_b/2)(x-x_b)^2$. We identify the position $x^* = (\kappa_bx_b + \kappa_lx_l)/(\kappa_l+\kappa_b)$ where the two approximations take a common slope and match the functions smoothly there. The resulting piecewise composite curve has a total rise of 
\begin{multline} 
U_{0,l}(x^*) - U_0(x_l) + U_{0,b}(x^*) - U_0(x_b)\\
= \frac{\kappa_l\kappa_b(x_b-x_l)^2}{2(\kappa_l+\kappa_b)}
= \frac{1}{2}\kappa^\ddagger(x^\ddagger)^2,
\end{multline}
which differs from the the true barrier height $\Delta G^\ddagger = U_0(x_b) - U_0(x_l)$ by a factor that Dudko labels $1/\nu$. That is,
\begin{equation} \label{EQ:shape-param-defn}
\frac{\Delta G^\ddagger}{\nu} = \frac{1}{2}\kappa^\ddagger(x^\ddagger)^2.
\end{equation}
The equality $\nu = 2/3$ holds for any degree-three polynomial. If the energy landscape is represented by a higher-degree polynomial, then the value of the shape parameter is idiosyncratic and should be viewed as drawn from a distribution with average $\langle 1/\nu \rangle < 3/2$. For smooth potentials (no cusps or discontinuities), typical values of the shape parameter $\nu$ range between $2/3$ and $\approx 1.1$. An advantage of working in terms of $\nu$, rather than the effective curvature $\kappa^\ddagger$, is that the former can be defined even if the derivatives $U''(x_l)$ and $U''(x_b)$ vanish (e.g., a quartic well or barrier) or are not well defined (e.g., a cusp barrier).

With Eq.~\eqref{EQ:shape-param-defn} in mind, matching our resummed rate expression to that of Dudko order by order in the small-pulling-force, large-barrier-height limit suggests the form
\begin{equation} \label{EQ:rate-renorm}
k(F) = k_0 \exp\Biggl[\frac{\beta Fx^{\ddagger}}{(1+ \alpha/\beta \Delta G^\ddagger)(1+ \nu Fx^{\ddagger}/4{\Delta G}^\ddagger)}\Biggr],
\end{equation}
where $\alpha > 0$ is a pure number with a weak dependence on the shape parameter.
Equation~\eqref{EQ:rate-renorm} can be understood as a rewriting of Eq.~\eqref{EQ:Bell-Evans}, the Bell-Evans phenomenological rate, with an upward renormalization of the temperature, $\beta \to {\beta}/(1+ \alpha/{\beta {\Delta G}^\ddagger})$, and a downward renormalization of the barrier distance $x^\ddagger \to x^\ddagger/(1+ \nu Fx^{\ddagger}/4{\Delta G}^\ddagger)$. Unlike Eq.~\eqref{EQ:rate-Dudko}, Eq.~\eqref{EQ:rate-renorm} is well behaved everywhere.

In the case of an ultrahigh barrier, defined by the double limit $\beta \Delta G^\ddagger \gg 1$ and $\Delta G^\ddagger \gg Fx^\ddagger$, Eq.~\eqref{EQ:rate-renorm} reduces to Eq.~\eqref{EQ:Bell-Evans}. For more modest barriers or higher temperatures, one or both of the factors $(1+ \alpha/\beta \Delta G^\ddagger)$ and $(1+ \nu Fx^{\ddagger}/4{\Delta G}^\ddagger)$ may differ appreciably from 1; this allows the rate expression to become aware of the details of the barrier's height and shape through the factor $\Delta G^\ddagger/\nu$.

The reliability of Eq.~\eqref{EQ:rate-renorm} was tested for six potential landscapes with different values of $\nu$ using the simulation scheme described in the next section. In every test example (see Fig.~\ref{FIG:nu}), our renormalized equation closely tracked the empirical escape rate determined from simulations. It noticeably outperformed the Bell-Evans and Dudko escape rate equations.

\newpage
\section{ \label{SEC:Numerical-Simulations}Numerical Simulations}

The reaction coordinate $x$ was made to execute Langevin dynamics according to
\begin{equation} \label{EQ:Langevin-Equation}
 m \ddot {x}= m\dot{\varv}  = -\pdv {U}{x} - \gamma \varv + \xi (t).
\end{equation}
This was implemented using a modern reformulation~\cite{Gronbech-Jensen-Mol-13} of the Verlet algorithm~\cite{Verlet-PRL-67}. We mimicked the experimental situation by assuming stochastic motion of a molecule of effective mass $m = \SI{2}{pg}$ in a biquadratic potential. The data that appear in Figs.~\ref{FIG:prob-distr}--\ref{FIG:kappa-dagger} correspond to the choice $U_0(x) = 4x^4 - 32x^2 +64$ (with $x$ measured in nm and $U_0$ in \si{pN.nm}). The molecule was assumed to be  pulled from two ends along the reaction coordinate $x$ by a laser potential with force constant $K$ and pulling velocity $V$; i.e., with an instantaneous force $F = KVt$ that increases linearly in time. For the given potential, the energy barrier was ${\Delta G}^\ddagger = \SI{64}{\pico\newton\nano\metre}$, the minimum-to-barrier distance $x^\ddagger = \SI{2}{nm}$, and the effective curvature $\kappa^\dagger = \SI{42.7}{\pico\newton\per\nano\metre}$. The stochastic forces $\xi (t) $ on the molecule were drawn randomly from a Gaussian distribution of width $(2m\gamma k_BT \delta t)^{1/2}$ with  $k_BT = \SI{4.1}{\pico\newton\nano\metre}$, $\gamma  = \valunit{7}{\per\us}$, and a discrete timestep $\delta t$ ranging from $\valunit{e-2}{\us}$ to $\valunit{e-6}{\us}$.

Note that, for generality, small inertial effects were included in the numerics. The simulations were not run in the strongly overdamped, diffusion-only limit: parameter values were chosen to be physically plausible but also to produce a nonextreme limit (neither $\gamma \ll \omega_b$ nor $\gamma \gg \omega_b$) of the prefactor to the exponential in the Kramers rate [which will appear in Eq.~\eqref{EQ:kramer's equation}].

Pulling rates for the force $F = KVt$ are measured with respect to $KV_\text{min} = k_0/\beta x^\ddagger$, which is the minimum rate for effectual pulling. Below $KV_\text{min}$, the probability density $p(F_c)$ is peaked at $F_c = 0$; the particle escapes the well on its own before the applied force has appreciably modified the energy landscape. On the other hand, for rates above $KV/KV_\text{min} \approx 10^6$, the barrier vanishes too quickly, long before the particle has moved any significant distance. Accordingly, we worked in the regime of pulling rates between these two extremes.

The simulation was initialized in the left well by drawing starting values of velocity $\varv$ and position $x$ from the distributions $e^{-\beta m \varv^2/2} $ and $e^{-\beta U(x)}\Theta(x_b-x)$, respectively, so that the each simulation began fully thermalized. The simulation flagged the value of pulling force at which $x$ convincingly crossed the barrier or the barrier vanished; we took this to be the rupture or critical force $F_c$. For each value of the pulling rate $KV$, the simulation was carried out 2500 times, each run generating a unique value of the rupture force. The cumulative probability distribution $P(F_c)$ was constructed in the standard way---by sorting the measured rupture forces in ascending order and then pairing them with a uniform grid of values running from zero to $1$.  The plot for $P(F_c)$ so obtained was tested against  Eq.~\eqref{EQ:cum-prob-resummed} and against the Bell-Evans form, Eq.~\eqref{EQ:cum-prob-BE}. The process was repeated for pulling rates ranging from $KV = \SI{e-7}{}$ to $\SI{0.6}{\pico\newton\per\us}$ (roughly $1 \lesssim KV/KV_\text{min} \lesssim 10^7$) to determine how these expressions fare in the slow, intermediate, and fast pulling regimes.

In order to test parameter extraction, the original $P(F_c)$ data set for each pulling rate was bootstrapped~\cite{Efron-93} 100 times to generate 100 new instantiations. These data were fitted with Eq.~\eqref{EQ:cum-prob-resummed} to extract the intrinsic parameters of the potential landscape: $k_0$, $x^\ddagger$, and $\kappa^\ddagger$. The spread in fit values was used to generate error estimates.

The data sets were also fitted to the Bell-Evans form given by Eq.~\eqref{EQ:cum-prob-BE} in order to extract the values of $k_0$ and $ x^\ddagger $ ($\kappa ^\ddagger$ does not appear in the Bell-Evans expression). The bootstrap-average   values of the extracted parameters were compared to their known values. The theoretical intrinsic rate $k_0$ was  computed according to the usual Kramers result,
\begin{equation}\label{EQ:kramer's equation}
k_0 = \frac{\omega_l}{2\pi} 
\frac{\sqrt{\gamma^2/4 + \omega _b^2} - \gamma/2}{\omega_b} 
\exp\Bigl(-\beta \Delta G^\ddagger\Bigr).
\end{equation}  
Our test potential corresponded to  $\omega_l = \sqrt{\kappa_l/m} = \valunit{8}{\per\us}$ and $\omega_b = \sqrt{\kappa_b/m} = \valunit{5.65}{\per\us}$. We verified that the theoretical value of $k_0 = \valunit{1.192e-7}{\per\us}$ was in agreement with numerical measurements of the escape rate for the nontilted energy landscape.

\section{ \label{SEC: Results and Conclusions} Results and Conclusions}

\begin{figure}
\begin{center}
\includegraphics[width=1.68in]{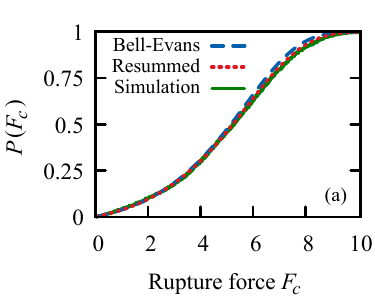}
\includegraphics[width=1.68in]{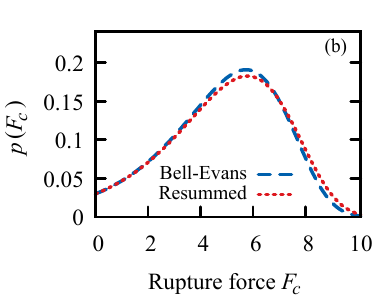}
\includegraphics[width=1.68in]{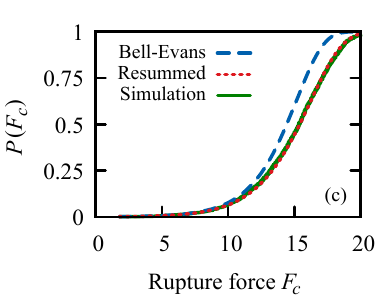}
\includegraphics[width=1.68in]{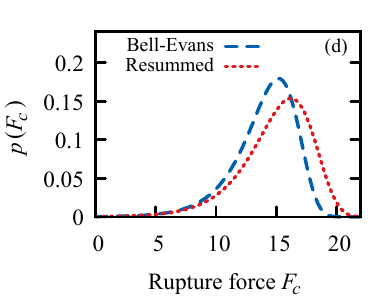}
\includegraphics[width=1.68in]{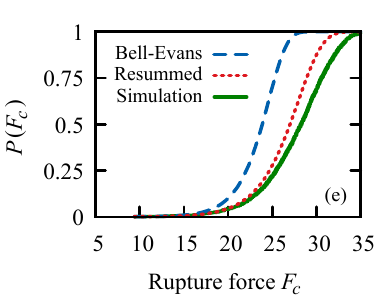}
\includegraphics[width=1.68in]{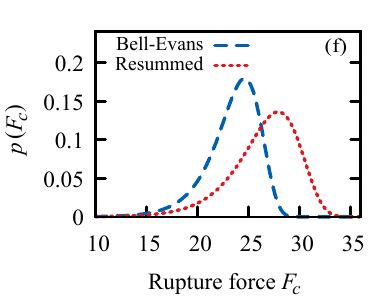}
\end{center}
\caption{\label{FIG:prob-distr}
Plots paired on the left and right show the cumulative probability distribution $P(F_c)$ and the corresponding probability density $p(F_c) = P'(F_c)$ of the rupture force. Each row shows results for successively faster pulling rates: (a,b) $KV = \SI{4e-6}{\pico\newton\per\us}$, (c,d) $KV = \SI{4e-4}{\pico\newton\per\us}$, and (e,f) $KV = \SI{4e-2}{\pico\newton\per\us}$.}
\end{figure}

In the left column of Fig.~\ref{FIG:prob-distr}, the rupture force distributions predicted by Eqs.~\eqref{EQ:cum-prob-BE} and \eqref{EQ:cum-prob-resummed} are compared to the results from simulation for three different pulling rates (corresponding to $KV/KV_\text{min} \approx 10^1$, $10^3$, and $10^5$). For slow pulling (top row), the  Bell-Evans theory and our resummed expression are well matched to each other and to the numerics. For intermediate pulling (middle row), the Bell-Evans result begins to deviate significantly, whereas our proposal continues to give accurate results (i.e., the solid green and red dotted lines coincide). Only at the highest pulling rates (bottom row) do we find significant deviation from the simulated rupture force distributions for both Eqs.~\eqref{EQ:cum-prob-BE} and \eqref{EQ:cum-prob-resummed}; although, even there, our expression performs better and is in good agreement up to $\sim \valunit{25}{pN}$.

\begin{figure}
\begin{center}
\includegraphics[width=2.75in]{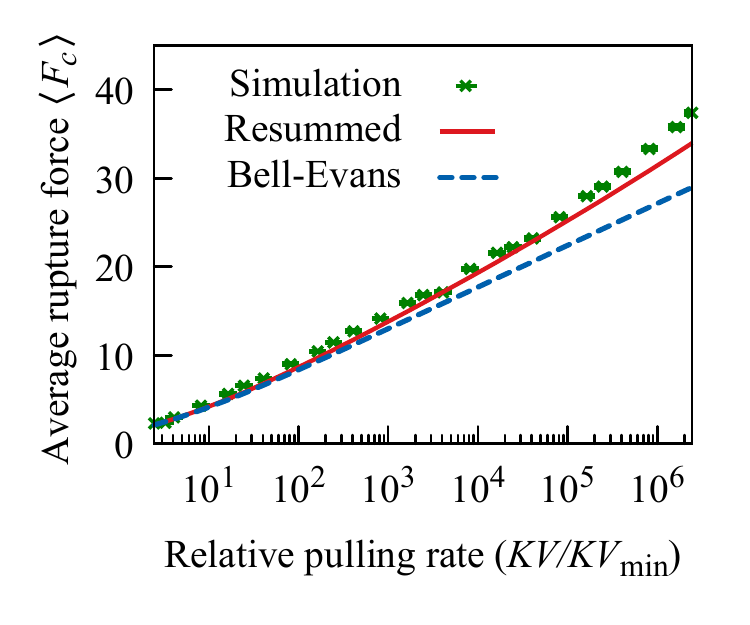}
\end{center}
\caption{\label{FIG:Fc-avg} 
The average rupture force $\langle F_c \rangle = \int\!dF\,FP'(F)$, determined by numerical integration  
with $P(F)$ taken from Eqs.~\eqref{EQ:cum-prob-BE} (dashed blue line) and \eqref{EQ:cum-prob-resummed} (red line),
is compared to the empirical values from simulation (green crosses).
}
\end{figure}

It is instructive to look at the corresponding probability density of the rupture force, $p(F_c) = P'(F_c)$, obtained from Bell-Evans and our resummed form, as shown in the right column of Fig.~\ref{FIG:prob-distr}. The Bell-Evans result systematically underestimates the pulling force required to traverse the barrier---and increasingly so for faster pulling. One observes that both its peak (typical rupture force) and its overall weight (mean rupture force) are positioned too far to the left (toward low force values). The same information is contained in the average critical force $\langle F_c \rangle$, which we obtained from the cumulative probability distributions, Eqs.~\eqref{EQ:cum-prob-BE}  and \eqref{EQ:cum-prob-resummed}, by numerical integration. Figure~\ref{FIG:Fc-avg} shows a plot of $\langle F_c \rangle$ as a function of the relative pulling rate. One can readily identify an intermediate regime ($10^2 \lesssim KV/KV_\text{min} \lesssim 10^5$) in which the curve computed from the resummed rate tracks the true, numerically determined values of the average rupture force. In that same regime, the Bell-Evans curve deviates significantly.

The second part of the numerical analysis focused on the inverse problem. Here, the simulated data were fitted using Eq.~\eqref{EQ:cum-prob-resummed}, and the intrinsic parameters of the energy landscape, viz., $k_0$, $x^\ddagger$, and $\kappa ^\ddagger$, were determined by minimizing discrepancies between theory and data in the least-squares sense. The process was repeated for Eq.~\eqref{EQ:cum-prob-BE}, but only with $k_0$ and $x^\ddagger$ (since $\kappa ^\ddagger$ does not appear in the fitting function). We found unambiguously that the parameter extraction is much more reliable using our resummed form. Indeed, use of the Bell-Evans theory was often quite misleading, because it would produce an apparently good fit that corresponded to incorrect values of the landscape parameters.

\begin{figure}[h!]
\begin{center}
\includegraphics[width=2.75in]{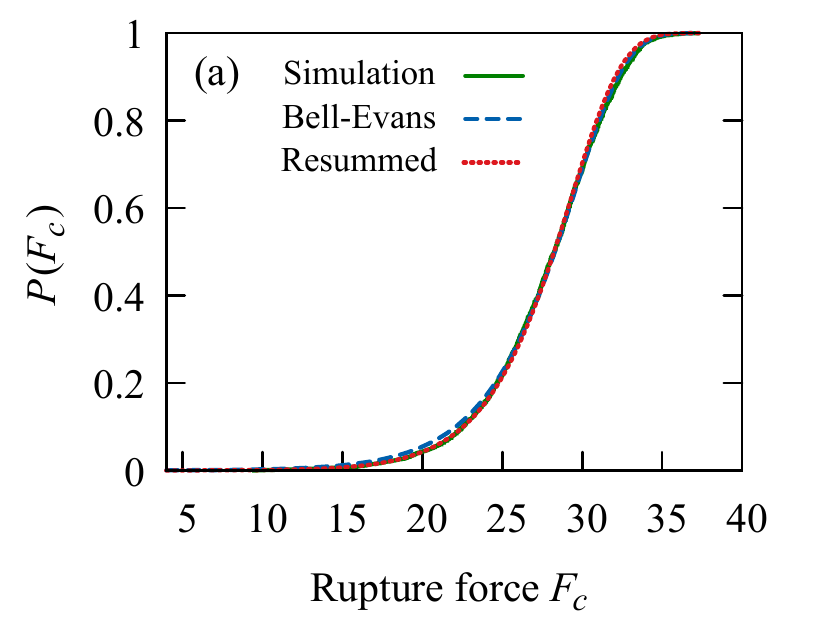}\\[-0.2cm]
\includegraphics[width=2.75in]{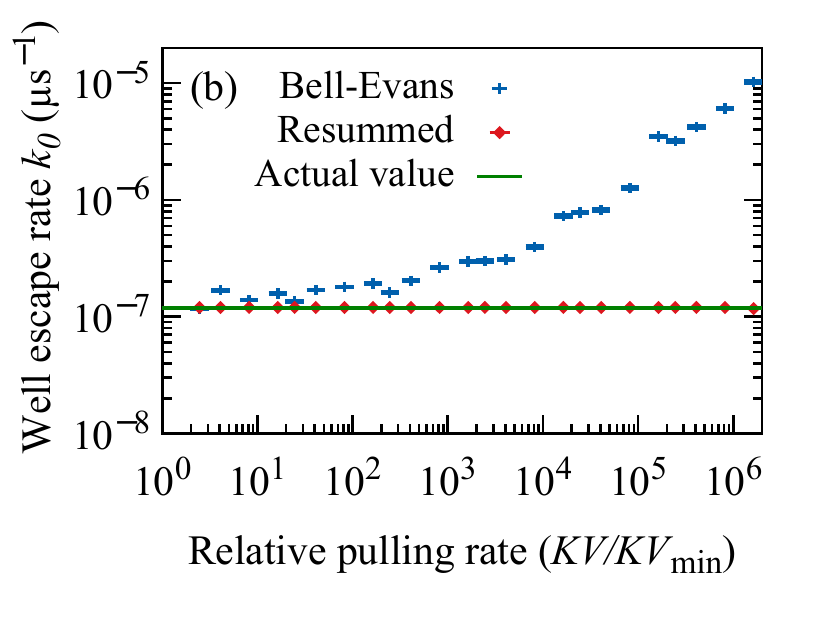}\\[-0.2cm]
\includegraphics[width=2.75in]{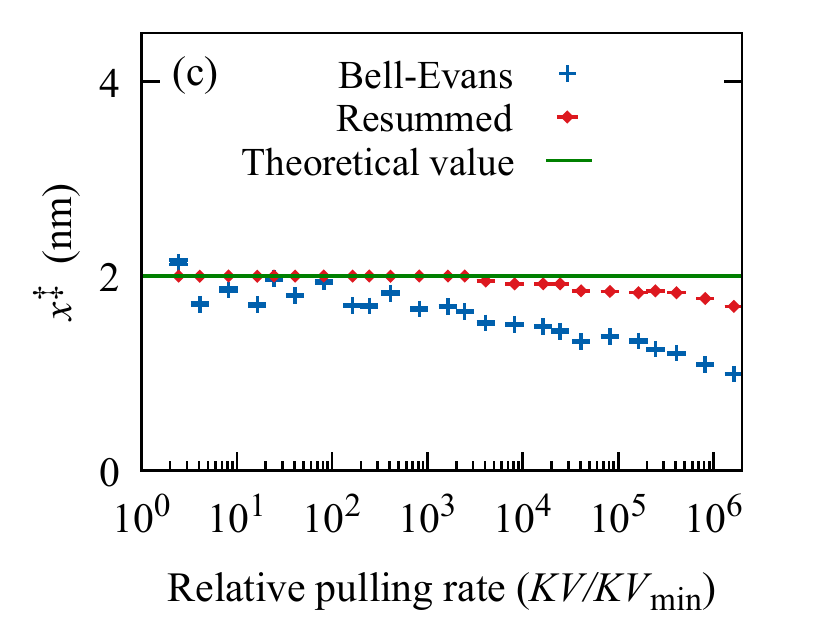}
\end{center}
\caption{\label{FIG:prob-k0-xdagger}
(a) The cumulative probability distribution of the rupture force, computed from 2500 simulated pulling experiments at rate $KV = \valunit{0.04}{pN\per\us}$ (green line), is plotted alongside the best fits for Eqs.~\eqref{EQ:cum-prob-BE} (blue dashed line) and \eqref{EQ:cum-prob-resummed} (red dotted line). The near indistinguishability of the curves illustrates the strong tendency toward overfitting. The Bell-Evans expression, though ill suited for describing the behavior at this high pulling rate, is able to mimic the numerical data---but at the cost of producing fitting parameters that have drifted far from their true values. This is in contrast to the poor agreement in Fig.~\ref{FIG:prob-distr}(e), where there is no fitting and the known values of $k_0$ and $x^\ddagger$ are used. Estimates of (b) the intrinsic escape rate $k_0$, and (c) the barrier distance $x^\ddagger$, as determined from fits of Eqs.~\eqref{EQ:cum-prob-BE} (blue crosses) and \eqref{EQ:cum-prob-resummed} (red diamonds) to simulation data over a range of pulling rates, are plotted alongside the actual value (green line).
}
\end{figure}

\begin{figure}
\begin{center}
\includegraphics[width=3.0in]{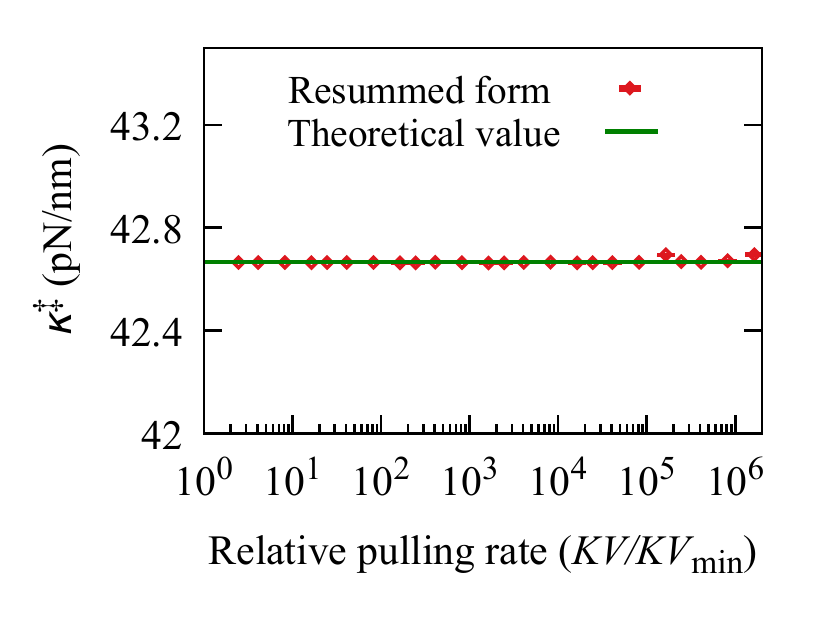}
\end{center}
\caption{\label{FIG:kappa-dagger} 
The plotted points (red diamonds) are estimates of the reduced curvature $\kappa^\ddagger$, as determined from fits of Eq.~\eqref{EQ:cum-prob-resummed} to simulation data. They compare favorably to the actual value (green line) over a range pulling rates spanning many decades.}
\end{figure}

The top panel of Fig.~\ref{FIG:prob-k0-xdagger}, which shows a fast-pulling example with $KV = \SI{4e-2}{\pico\newton\per\us}$, emphasizes this point. The cumulative probability distribution appears to be equally well fit by Eqs.~\eqref{EQ:cum-prob-BE} and Eq.~\eqref{EQ:cum-prob-resummed}. The middle and bottom panels reveal this to be illusory. In the Bell-Evans analysis, the value of $k_0$ is systematically overestimated and $x^\ddagger$ underestimated, and both ever more so as the pulling rate is ramped up. On the other hand, the analysis based on our resummed form yields values consistent
with the correct landscape parameters. Moreover, even at low pulling rates, where Bell-Evans performs not too badly, our proposal is more reliable and produces less scatter in the parameter values.

We remark that fits of the simulation data to Eq.~\eqref{EQ:cum-prob-resummed} yield astonishingly good values of $\kappa^\ddagger$, the effective curvature (see Fig.~\ref{FIG:kappa-dagger}). In almost every case, regardless of pulling rate, the predicted value of $\kappa^\ddagger $ coincides with the true value. This suggests to us that our inclusion of higher-order corrections in the rate equation plays an important role in improving the overall quality of the parameter extraction.

To conclude, our work highlights the known inadequacies of the Bell-Evans phenomenological well escape rate. It also suggests that the celebrated equation due to Dudko and co-workers is not an adequate fix. We propose a new expression, Eq.~\eqref{EQ:rate-resummed}, that improves on the Bell-Evans expression by including beyond-Arrhenius contributions from the shape of the energy potential. Equation~\eqref{EQ:rate-resummed} clearly outperforms the Bell-Evans and Dudko expressions in terms of predicting the well escape rate (as is evident from Fig.~\ref{FIG:nu}). Crucially, it avoids the unphysical behavior that plagues Dudko's rate equation at large pulling force. 

Of particular utility is that Eq.~\eqref{EQ:rate-resummed} integrates to give a manageable, closed-form expression for the cumulative probability distribution. The resulting Eq.~\eqref{EQ:cum-prob-resummed} is straightforward to implement as a fitting function and can be incorporated into existing  workflows with little additional effort. 
Rigorous numerical tests (illustrated in Figs.~\ref{FIG:prob-k0-xdagger} and \ref{FIG:kappa-dagger}) confirm that fits to Eq.~\eqref{EQ:cum-prob-resummed} can be used to reliably extract the parameters that characterize the underlying energy landscape.

\acknowledgments

One of the authors (SA) thanks Thomas Perkins (JILA, NIST, and the University of Colorado Boulder) for helpful discussions.

\end{document}